# Automatic fault detection on BIPV systems without solar irradiation data

Jonathan Leloux[1,2,*], Luis Narvarte[1,2], Alberto Luna[1,2], Adrien Desportes[3]
[1]Instituto de Energía Solar – Universidad Politécnica de Madrid (IES-UPM), Madrid, Spain
[2]WebPV, Madrid, Spain
[3]Rtone, Lyon. France

[*]Corresponding author: jonathan.leloux@ies-def.upm.es

ABSTRACT: BIPV systems are small PV generation units spread out over the territory, and whose characteristics are very diverse. This makes difficult a cost-effective procedure for monitoring, fault detection, performance analyses, operation and maintenance. As a result, many problems affecting BIPV systems go undetected. In order to carry out effective automatic fault detection procedures, we need a performance indicator that is reliable and that can be applied on many PV systems at a very low cost. The existing approaches for analyzing the performance of PV systems are often based on the Performance Ratio (PR), whose accuracy depends on good solar irradiation data, which in turn can be very difficult to obtain or cost-prohibitive for the BIPV owner. We present an alternative fault detection procedure based on a performance indicator that can be constructed on the sole basis of the energy production data measured at the BIPV systems. This procedure does not require the input of operating conditions data, such as solar irradiation, air temperature, or wind speed. The performance indicator, called Performance to Peers (P2P), is constructed from spatial and temporal correlations between the energy output of neighboring and similar PV systems. This method was developed from the analysis of the energy production data of approximately 10,000 BIPV systems located in Europe. The results of our procedure are illustrated on the hourly, daily and monthly data monitored during one year at one BIPV system located in the South of Belgium. Our results confirm that it is possible to carry out automatic fault detection procedures without solar irradiation data. P2P proves to be more stable than PR most of the time, and thus constitutes a more reliable performance indicator for fault detection procedures. We also discuss the main limitations of this novel methodology, and we suggest several future lines of research that seem promising to improve on these procedures.

1 INTRODUCTION

Photovoltaic (PV) systems can be classified into two main groups:
- Large ground-mounted PV plants, whose nameplate power stands somewhere between 1 MW and 1 GW.
- Building Integrated Photovoltaics (BIPV).

The Operation and Maintenance (O&M) of PV plants is commonly assisted by Supervisory Control And Data Acquisition (SCADA), which register sizeable quantities of information relating to the operation of the system. These data often include power, voltage, current, and energy output measured up to the PV module string level, at Direct Current (DC) and Alternating Current (AC) levels. A weather station also measures solar irradiation, air temperature and wind speed [1,2,3].

BIPV systems bear important differences respect to PV plants:
- They consist of a large number of small PV generation units, whose power is usually of some kW, spread out over the territory.
- They differ from one another by their size, components, orientation and tilt angles, topology, and quality.
- They operate under weather conditions that vary from one installation to another, both in space and in time.
- They usually each belong to a different owner.
- Their energy output recorded by the energy meter is often the only accurate information available from them.

This reality makes difficult a cost-effective procedure for monitoring, fault detection, performance analyses, operation and maintenance. As a result, many problems affecting BIPV systems go undetected.

Previous works have reported and analyzed the performance of thousands of BIPV systems worldwide [4,5,6,7,8,9,10]. A recent review of the performance of 10,000 residential PV systems in France and Belgium concluded that, on average, the performance of a PV system is 15% lower than what could be achieved [11,12]. These studies have identified and quantified the main causes explaining the performance losses, and they have drawn a general picture of the state of the art.

BIPV performance diagnosis solutions are needed:
- BIPV owners show interest in the follow-up of their system's performance [13,14,15].
- BIPV installers are looking for efficient performance analysis procedures [16].
- Policy makers need more data to properly target the most relevant challenges [17,18].
- Science and industry require feedback from the field to pursue technological improvement further [19,20,21,22,23,24].

The existing approaches for analyzing the performance of PV systems are often based on ratios between the energy output of a PV system and the solar irradiation that it receives [25,26,27].

The *Performance Ratio* (*PR*) is, by far, the most commonly used performance indicator today. It is defined as:

$$PR = \frac{E_{PV}}{\frac{P^*}{G^*}\int_T G\,dt}$$

where:
- $T$ is the time interval during which the solar irradiation was received and the PV energy was produced.
- $E_{PV}$ is the electrical energy output produced by the PV system.
- \* refers to Standard Test Conditions (STC).
- $P^*$ is the PV array's rated power under STC.
- $G^*$ is the global solar irradiance under STC (i.e. $G^* = 1000$ W/m$^2$).
- $G$ is the global solar irradiance received by the PV generator.

In order to perform an effective automatic fault detection procedure for BIPV, we need a performance indicator whose main characteristics are:
- It is as stable as possible in absence of failure.
- It drops significantly in presence of a failure.
- It can be applied at a low cost.

Figure 1 shows the evolution of hourly *PR* of a BIPV system in Belgium during one year. It shows that *PR* is not stable enough to allow for accurate fault detections.

Several causes can produce uncertainties in *PR*:
- It is difficult to obtain reliable data on *G*. The data measured by pyranometers are not available at a sufficient spatial resolution. The data retrieved from satellites can convey important inaccuracies, due in particular to cloudiness, snow cover, aerosols, and low solar elevations angles.
- Thermal losses affecting PV systems depend on the weather conditions, notably including air temperature, and wind speed [28].

Additionally, accurate solar irradiation, air temperature and wind speed data can be cost-prohibitive for the BIPV owner.

In consequence, the use of *PR* for automatic fault detection procedures for BIPV faces several weaknesses.

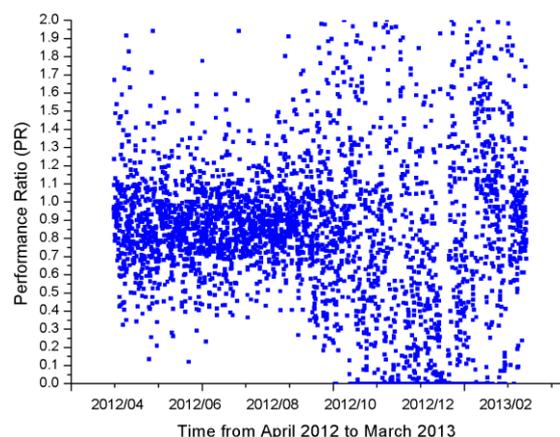

**Figure 1:** Hourly *PR* of a BIPV system during one year. *PR* shows wide variations, which makes its use more difficult as a fault detector for BIPV.

We present an alternative fault detection procedure. It is based on a novel performance indicator that can be constructed on the sole basis of the energy production data measured at the BIPV systems. This procedure does not require the input of operating conditions data, such as solar irradiation, air temperature, or wind speed.

The performance indicator is constructed from spatial and temporal correlations between the energy output of neighboring BIPV systems. The resulting performance indicator has been designated as Performance to Peers (*P2P*), because it is based on comparisons between neighboring and similar installations, i.e. *peer* PV systems.

In this contribution, we explain the main steps leading to the construction of this novel performance indicator. We then explain how we use this performance indicator to perform automatic fault detection procedures. We also show the results of applying these procedures on one BIPV system.

2 DATA

The energy production data were collected by Rtone on approximately 10,000 BIPV systems located in Europe, mainly in Belgium, France and UK. These data correspond to approximately 3 years of operation (2011-2014) and were measured with a temporal resolution of 1 data each 10 minutes. These data where obtained through energy meters equipped with General Packet Radio Service (GPRS). Figure 2 shows the location of these BIPV systems monitored by Rtone in Europe.

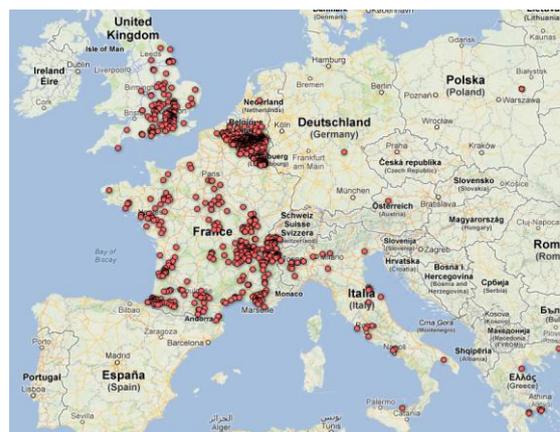

**Figure 2:** Location of the 10,000 BIPV systems monitored by Rtone in Europe. These systems are mainly located in France, Belgium and UK.

Most of the BIPV systems that were analyzed in the context of this work are residential Building Added PV (BAPV) installations of a peak power from 1 to 10 kW, and located in the South of Belgium (Wallonia). Figure 3 shows that the density of the PV installations monitored by Rtone is very high in this region. This context provides an ideal research playfield for developing a performance indicator based on comparisons between the energy outputs of neighboring installations.

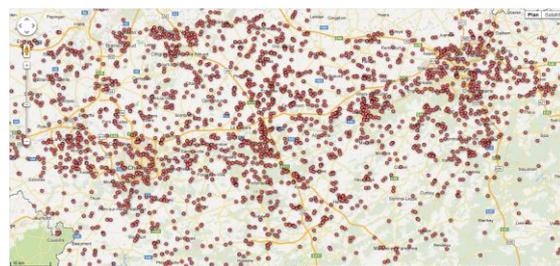

**Figure 3:** BIPV installations monitored by Rtone in the South of Belgium. In this region, the density of the BIPV installations monitored by Rtone is very high.

The results of our procedure are illustrated on the hourly, daily and monthly data monitored at one BIPV system located in the South of Belgium, and corresponding to one year of operation (April 2012 – March 2013).

## 3 METHODOLOGY

Our automatic fault detection procedure consists of two main consecutive steps:
1) Construct a performance indicator from a comparison of the energy outputs of neighboring PV systems. This performance indicator needs to be as stable as possible when the PV system is not affected by performance problems.
2) Detect a performance problem from an abnormal variation in this performance indicator.

Each one of these two steps is explained in this section. This procedure was elaborated and patented by IES-UPM [29,30].

### 3.1 Construction of the Performance to Peers (P2P)

The P2P is constructed from comparison between the energy produced by a PV system and the energy produced by its peers. This is in turn achieved through three main successive steps:
A) Normalize the energy outputs from all the PV systems. This makes comparisons easier.
B) Determine and quantify the degree of correlation that exists between the energy outputs of any two PV systems.
C) Calculate P2P. This is achieved from comparison of the normalized energy outputs of a given PV system and of its neighbors. The comparison between two neighboring PV systems is given more weight if their degree of correlation is higher.

The next paragraphs describe these three steps.

A) Normalize the energy outputs from PV systems

The Capacity Utilization Factor (*CUF*) is defined as:

$$CUF = \frac{E_{PV}}{P^* T}$$

*CUF* allows normalizing the energy output of a PV system ($E_{PV}$) by its peak power ($P^*$) and the time interval (*T*) during which the energy was produced. It thus allows comparing the energy output corresponding to PV systems of a different peak power, and during different time intervals. *CUF* is expressed in [%] and it represents the fraction of the energy produced by the PV system relatively to the energy that would have been produced during the same time interval if the PV system were producing at peak power during the whole time interval. A *CUF* of 100% means that the system has produced energy equivalent to $P^*$ times *T* during the time interval *T*.

B) Quantify the correlation between the PV systems

Among all the possible neighboring PV systems available, we need to determine which ones are the best peers for a given PV system.

The neighboring PV systems usually differ in PV technology, orientation and tilt angles, PV array's peak power, and components' quality, and even the information provided by BIPV owners and installers on these characteristics is often inaccurate [31].

Furthermore, neighboring PV systems can often operate under different weather conditions, in particular when considering short time intervals.

The only accurate information that is available to us is often the energy output data. Therefore, we have developed an algorithm that is able to identify the best peer PV systems from the degree of correlation between their *CUF*. For that purpose, we use statistical correlations based on the variance and on the covariance of the ratio between the *CUF* of two PV systems.

C) Calculate the Performance to Peers (P2P)

The *P2P* of a given system is calculated from comparison between its *CUF* and the *CUF* of its peers. This comparison between two peer PV systems is given more weight if the degree of correlation of their *CUF* is higher.

### 3.2 Automatic fault detection from P2P variations

The automatic fault detection procedure is carried out through the establishment of a minimum threshold value of *P2P*, below which the performance of the PV system is considered as abnormally low.

Figure 4 shows the histogram of daily *P2P* corresponding to one BIPV installation in Belgium for each day of the year 2012. We can identify two distinct distributions.

- Most of the *P2P* values (in green) are clustered at the right of the graph, around values close to 1. These values are distributed following a somewhat normal (Gaussian) distribution. These *P2P* values are representative of the normal (faultless) functioning of the PV system. The Gaussian distribution itself is due to the intrinsic uncertainties affecting *P2P*.
- Other *P2P* values (in red), much less frequent, are found anywhere between 0 and 0.9. These *P2P* values correspond to performance problems (or faults).

The fault detection threshold needs to separate these two different *P2P* populations. The human eye would set this threshold around *P2P* values of 0.93. In that region (in orange), there is unavoidable overlap between the two populations. This implies that for these intermediate *P2P* values, it is impossible to know for certain whether they belong to normal or abnormal operation. Our fault detector will therefore qualify some faulty *P2P* values as normal, and some normal *P2P* values as faulty. In statistical hypothesis testing [32], these kinds of errors of misjudgments are called errors of type I and type II. These errors are incorrect rejection of, respectively, a true null hypothesis, and incorrect failure to reject a false null hypothesis. More simply stated, a type I error is detecting a performance problem that is not present, while a type II error is failing to detect a performance problem that is present. The exact value of the threshold is therefore a matter of compromise between the two types of errors.

We have chosen to establish the threshold value so that the probability of error of type I (i.e., to state the existence of a performance problem when the PV system is functioning properly, or put in another way, to emit a false alarm) is below 0.3%.

We find this threshold as follows:
- We locate the center of the Gaussian distribution. We do it by calculating the median of the whole distribution.
- We calculate the standard deviation of the Gaussian distribution. Before calculating this standard deviation, we need to exclude the abnormal *P2P* values, which would increase it artificially. We calculate standard deviation of these normal functioning values by calculating the standard deviation of the *P2P* values located at the right of the central value of the Gaussian distribution.
- We establish the value of the threshold as the central value of the Gaussian less three times the standard deviation of this Gaussian distribution. This leads to a probability of type I error of 0.3% [32].

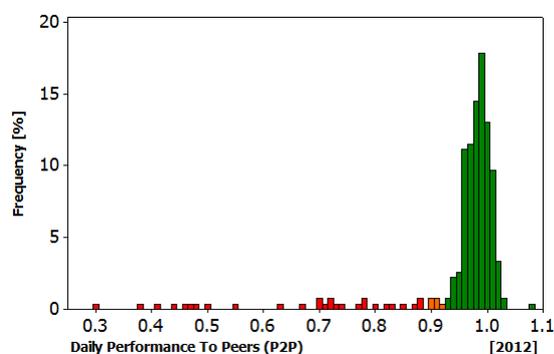

**Figure 4:** Histogram of daily *P2P* on a BIPV system during the whole year 2012. The green bars are representative of the normal operation of the PV system. The red bars are probably due to performance problems. The orange bars correspond to an intermediate zone where there is an overlap between the two populations.

4 RESULTS

We illustrate the application of our procedure to one BIPV system located in the South of Belgium, during a period of one year, from April 1st, 2012, to March 31st, 2013.

The resulting *P2P* calculated for this installation is presented at hourly, daily, and monthly levels.

The corresponding *PR* values have also been represented on this figure to allow for comparisons between both performance indicators.

The application of the automatic fault detector is shown on the hourly data.

4.1 Hourly P2P
Figure 5 shows the hourly *P2P* and *PR*. The *P2P* turns out to be much more stable than *PR* overall.

The period from December 2012 to March 2013 is characterized by *P2P* values notably low. This period was marked by episodes of snow, which covered partially or totally the PV generators. During this period, we therefore simultaneously observed low energy production and low *P2P* values.

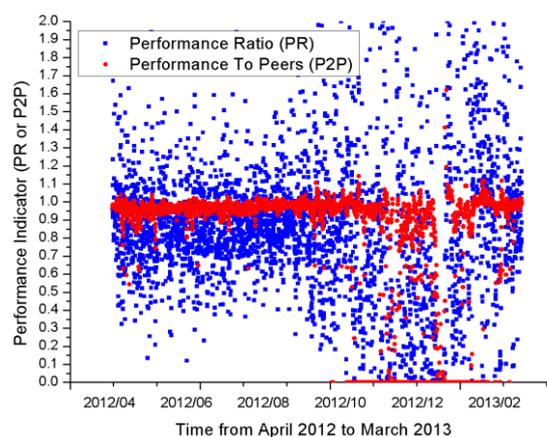

**Figure 5:** Hourly *P2P* and *PR* values observed on one BIPV system from April 1st, 2012 to March 31st, 2013.

Figure 6 shows the histogram corresponding to all the *P2P* values of figure 5. The histogram shows a Gaussian trend for the *P2P* values that correspond to a proper functioning of the PV system. Performance problems appear as *P2P* values that are significantly lower and are spread over a large range of values, corresponding to different kinds of performance problems. A group of zero-values of *P2P* is also visible on the left of the histogram, corresponding to zero-production faults. The failure detection threshold was calculated as 0.65.

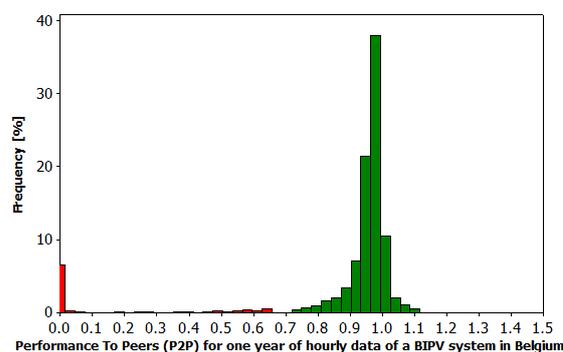

**Figure 6:** Histogram of hourly *P2P* and *PR* values observed on one BIPV system from April 1st, 2012 to March 31st, 2013.

3.2 Daily P2P
Figure 7 shows the daily *P2P* and *PR*. The *P2P* is relatively stable over the whole period, except during the period already mentioned before and corresponding to snow episodes, where *P2P* falls notably. This period is also marked by a high variability of *PR* values. Similar observations were made on the majority of the BIPV systems in Belgium, and were tracked back to high inaccuracy in the solar radiation data provided by the satellites in presence of snow cover, high cloudiness and low solar angles. The corresponding failure threshold was calculated to be 0.77. This value is higher than for hourly values, because the P2P is more stable at a daily level.

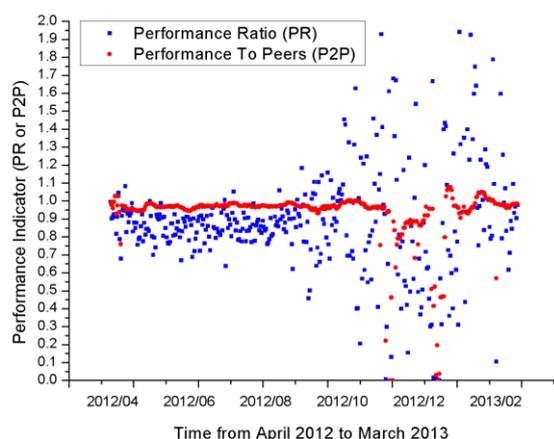

**Figure 7:** Daily *P2P* and *PR* values observed on one BIPV system from April 1st, 2012 to March 31st, 2013.

4.3 Monthly P2P

Figure 8 shows the monthly *P2P* and *PR*. Similarly to what was observed at hourly and daily levels, the *P2P* is stable over the whole period, showing lower values during the snow episode of the winter. The *PR* shows a seasonal trend due to the effect of temperature, with lower values during the summer months. This seasonal trend is not visible on *P2P*. The corresponding fault detection threshold was calculated as 0.775, which is nearly the same as for daily *P2P*.

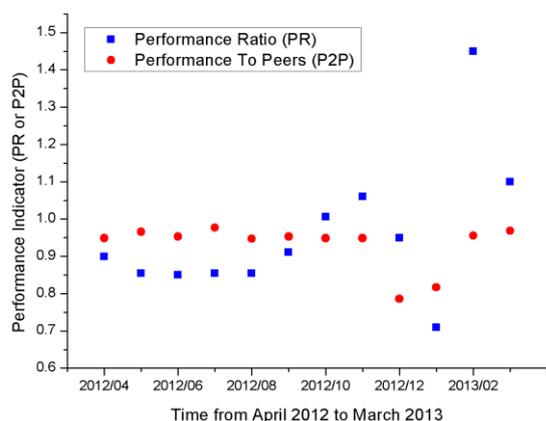

**Figure 8:** Monthly *P2P* and *PR* values observed on one particular BIPV system in Belgium from April 1st, 2012 to March 31st, 2013.

5 DISCUSSION

Our results confirm that it is possible to carry out automatic fault detection procedures without solar irradiation data.

*P2P* proves to be more stable than *PR* most of the time, and thus constitutes a more reliable performance indicator for fault detection procedures.

Even though the energy output data that we analyzed were monitored at a 10-min level, we found that it was generally unnecessary and even counterproductive to construct a *P2P* below the hourly level. The main reason is that the 10-min P2P are more unstable under fast changing weather conditions. The integration of 10-min data into hourly data plays the role of a low-pass filter which greatly improves the stability of *P2P*.

The goodness of *P2P* applied to one PV system depends on the number of peer PV installations located nearby. It is not straightforward to determine how many peer installations are necessary to lead to a *P2P* that is accurate enough to allow for effective fault detection procedures. Generally, we have come to the conclusion that most of the time, 10 peers at less than 10 kilometers lead to very good results. When fewer peers are available, then the accuracy of *P2P* depends on several parameters:

- The temporal resolution at which the fault detection is attempted. The higher is the temporal resolution, the higher is the variability of *P2P* due to the dynamical character of weather fluctuations. It is much easier to obtain a reliable *P2P* at daily level than at hourly level.
- The kind of performance problem that needs to be detected. Some faults induce large variations in *P2P*, while others are only visible through small variations of it. The faults that are more difficult to detect require a more accurate *P2P*. Fortunately, in general the faults that are the easiest to detect are also the most relevant in terms of energy losses.
- The similarity between the main characteristics of one PV installation (orientation and tilt angle, PV module technology…) respect to its neighbors. If a PV installation presents characteristics that are very uncommon, the energetic behavior of its neighbors presents a poorer correlation with it, making more difficult the construction of a reliable *P2P*. This can happen when the PV installation has few neighbors, has uncommon orientation and tilt, or is equipped with uncommon PV module technology.
- The weather conditions play an important role on the goodness of *P2P*. It is much easier to obtain a stable *P2P* when all the neighboring PV systems are under clear-sky conditions, than in presence of cloud covers composed of many small and fast moving cumulus that provoke very heterogeneous weather conditions at extremely local levels. Under these conditions, previous works [33] have demonstrated that two PV systems separated only one kilometer from each other can receive a solar irradiation that can greatly differ at hourly level, and even at daily level. This translates into a problem of spatial and temporal resolution. Fortunately, we do not always need to detect problems under any weather conditions, or at each moment. We can filter out these too unfavorable weather conditions and apply our procedures only to the more favorable conditions.

When not enough peers are available and *P2P* is not stable enough, then we use a fault detector that is constructed from solar irradiation data. For this purpose, we construct a Performance Index (*PI*) [11,12]. *PI* is a ratio between the energy produced by a PV system, and the energy that this system would produce if it were free of any kind of avoidable performance losses. A PI of 100% corresponds to a PV system equipped with an inverter and a PV generator whose real power and characteristics coincide with their rated nominal value, and which is free of failures, shading, soiling and wiring losses. *PI* still suffers from the uncertainties in solar irradiation, but it is more stable than *PR*, because it does not suffer from the variations due to the PV generator's thermal losses and the other avoidable energy losses.

The good results obtained from the use of *P2P* do not imply that solar irradiation data are useless. Solar irradiation data and *P2P* are no enemies, and they are even complementary.

Several researches are currently working on novel methods that are able to improve on the solar irradiation data provided by satellites by merging them with ground measurements carried out with pyranometers [34,35].

Other works have also demonstrated that the power output of a PV module can be used as a sensor of solar irradiance [36]. Our recent research activities have shown that solar irradiation data could also be obtained from the energy output data of neighboring PV systems [37].

There are thus interesting possibilities to improve on the quality of solar irradiation data by combining the information provided by satellites, pyranometers and PV installations.

This contribution has presented a method that allows to detect a performance problem, but not to identify which is its cause. This is nevertheless possible, and it is currently the subject of intense research from our part, and we have already been able to diagnose several of the most relevant performance problems. These performance diagnosis procedures are out of the scope of the present article.

These automatic fault detection procedures are now available for the PV sector through Web services that are commercialized by WebPV [38], a spin-off company from IES-UPM. These Web services are particularly suitable for BIPV installers or monitoring companies.

We thereby hope to collect as much data as possible, as diversified as possible, in order to continue the research, improve on our models, and offer more services to the PV community.

6 CONCLUSION

Our results confirm that it is possible to carry out automatic fault detection procedures without solar irradiation data.

*P2P* proves to be more stable than *PR* most of the time, and thus constitutes a more reliable performance indicator for fault detection procedures.

The goodness of *P2P* applied to one PV system depends on the number of peer PV installations located nearby.

There are interesting possibilities to improve on the quality of solar irradiation data and fault detection procedures by combining the information provided by satellites, pyranometers and PV installations.

These automatic fault detection procedures are now available for the PV sector through Web services. These Web services are particularly suitable for BIPV installers or monitoring companies.

ACKNOWLEDGMENTS

The authors are grateful Jesús Robledo and David Trebosc from WebPV, for their help on the implementation of the fault detection procedures on the Web server. Fabien Grenier from Rtone was very effective providing help on the Web services. Many thanks to Catherine Praile, who carried out a great proofreading job, as usual. We benefitted from fruitful discussions with Manuel Collares-Pereira. This work has been partially supported by European Commission within the project PV CROPS [39] (Photovoltaic Cost r€duction, Reliability, Operational performance, Predicition and Simulation) under the 7$^{th}$ Framework Program (Grant Agreement nº 308468).

REFERENCES

[1] A. Chouder, S. Silvestre, Automatic supervision and fault detection of PV systems based on power losses analysis, Energy Conversion and Management, 2010.
[2] S. Silvestre, A. Chouder, E. Karatepe, Automatic fault detection in grid connected PV systems, Solar Energy, 2013.
[3] F. Spertino, F. Corona, Monitoring and checking of performance in photovoltaic plants: A tool for design, installation and maintenance of grid-connected systems, Renewable Energy, 2013.
[4] T. Oozeki, T. Yamada, K. Otani, T. Takashima, K. Kato, An analysis of reliability in the early stages of photovoltaic systems in Japan, Progress in Photovoltaics, 2010.
[5] G. Colantuono, A. Everard, L. M.H. Hall, A.R. Buckley, Monitoring nationwide ensembles of PV generators: Limitations and uncertainties. The case of the UK, Solar Energy, 2014.
[6] U. Jahn, W. Nasse, Operational performance of grid-connected PV systems on buildings in Germany. Progress in Photovoltaics, 2004.
[7] Y. Ueda, K. Kurokawa, K. Kitamura, M. Yokota, K. Akanuma, H. Sugihara, Performance analysis of various system configurations on grid-connected residential PV systems, Solar Energy Materials and Solar Cells, 2009.
[8] H. Yang, G. Zheng, C. Lou, D. An, J. Burnett, Grid-connected building-integrated photovoltaics: a Hong Kong case study, Solar Energy, 2004.
[9] S.A. Omer, R. Wilson, S.B. Riffat, Monitoring results of two examples of building integrated PV (BIPV) systems in the UK, Renewable Energy, 2003.
[10] M.A. Eltawil, Z. Zhao, Grid-connected photovoltaic power systems: Technical and potential problems — A review, Renewable and Sustainable Energy Reviews, 2010.
[11] J. Leloux, L. Narvarte, D. Trebosc, Review of the performance of residential PV systems in Belgium, Renewable and Sustainable Energy Reviews, 2012.
[12] J. Leloux, L. Narvarte, D. Trebosc, Review of the performance of residential PV systems in France, Renewable and Sustainable Energy Reviews, 2012.
[13] Sonnenertrag (http://www.sonnenertrag.de/).
[14] BDPV (http://www.bdpv.fr).
[15] PVOutput (http://www.pvoutput.org/).
[16] T. Dinkel, T. Koerner, Solar monitoring is a differentiator to today's PV installer, Renewable Energy World, 2011.
[17] R. Wüstenhagen, M. Wolsink, M.J. Bürer, Social acceptance of renewable energy innovation: An introduction to the concept, Energy Policy, 2007.
[18] K.H. Solangi, M.R. Islam, R. Saidur, N.A. Rahim, H. Fayaz, A review on global solar energy policy, Renewable and Sustainable Energy Reviews, 2011.
[19] International Energy Agency, Photovoltaic Power Systems Programme, Task 13, Performance and Reliability of Photovoltaic Systems (http://iea-pvps.org/index.php?id=57).
[20] PV PERFORMANCE, EU FP6 project (http://www.pv-performance.org/).
[21] F. Cucchiella, I. D'Adamo, Estimation of the energetic and environmental impacts of a roof-mounted building-integrated photovoltaic systems, Renewable and Sustainable Energy Reviews, 2012.
[22] M. Castro, A. Delgado, F.J. Argul, A. Colmenar, F. Yeves, J. Peire, Grid-connected PV buildings: analysis of future scenarios with an example of Southern Spain, Solar Energy, 2005.

[23] P. Padey, D. Le Boulch, I. Blanc, From Detailed LCA to Simplified Model: an Oriented Decision Makers Approach to Assess Energy Pathways, 2013.
[24] M.J. de Wild-Scholten, Energy payback time and carbon footprint of commercial photovoltaic systems, Solar Energy Materials and Solar Cells, 2013.
[25] A. Drews, A.C. de Keizer, H.G. Beyer, E. Lorenz, J. Betcke, W.G.J.H.M. van Sark, W. Heydenreich, E. Wiemken, S. Stettler, P. Toggweiler, S. Bofinger, M. Schneider, G. Heilscher, D. Heinemann, Monitoring and remote failure detection of grid-connected PV systems based on satellite observations, Solar Energy, 2007.
[26] S. Stettler, P. Toggweiler, J. Remund, SPYCE: Satellite photovoltaic yield control and evaluation, 21st European Photovoltaic Solar Energy Conference, 2006.
[27] A. Woyte, M. Richter, D. Moser, S. Mau, N. Reich, U. Jahn, Monitoring of phovoltaic systems: good practices and systematic analysis, 28th European Photovoltaic Solar Energy Conference and Exhibition, 2014.
[28] L. Gaillard, S. Giroux-Julien, C. Ménézo, H. Pabiou, Experimental evaluation of a naturally ventilated PV double-skin building envelope in real operating conditions, Solar Energy, 2014.
[29] A. Luna, Monitorización y análisis de la productividad de instalaciones fotovoltaicas, Final Course Project, Universidad Politécnica de Madrid, 2013.
[30] J. Leloux, L. Narvarte, R. Moretón, E. Lorenzo, Método de detección automática y diagnóstico de fallos de operación en instalaciones solares fotovoltaicas distribuidas, basado en la comparación de sus producciones de energía, Spanish Patent Nº P201430369, 2014.
[31] K. Kiefer, N.H. Reich, D. Dirnberger, C. Reise, Quality assurance of large scale PV power plants, 37th IEEE Photovoltaic Specialists Conference (PVSC), 2011.
[32] D. Sheskin, Handbook of Parametric and Nonparametric Statistical Procedures, CRC Press, 2004.
[33] M. García, L. Marroyo, E. Lorenzo, J. Marcos, M. Pérez, Solar irradiation and PV module temperature dispersion at a large-scale PV plant, Progress in Photovoltaics, 2014.
[34] M. Journée, C. Bertrand, Improving the spatio-temporal distribution of surface solar radiation data by merging ground and satellite measurements, Remote Sensing of Environment, 2010.
[35] M. Journée, R. Müller, C. Bertrand, Solar resource assessment in the Benelux by merging Meteosat-derived climate data and ground measurements, Solar Energy, 2012.
[36] F. Martínez-Moreno, E. Lorenzo, J. Muñoz, R. Moretón, On the testing of large PV arrays, Progress in Photovoltaics, 2012.
[37] J. Leloux, L. Narvarte, R. Moretón, E. Lorenzo, Método de generación de datos de irradiación solar a partir de datos de producción energética de instalaciones solares fotovoltaicas, Spanish Patent Nº P201431297, 2014.
[38] WebPV (http://www.webpv.net/).
[39] PV CROPS (Photovoltaic Cost r€duction, Reliability, Operational performance, Predicition and Simulation), European Commission FP7 project (http://www.pvcrops.eu/).